\documentclass[%
 reprint,
 amsmath,amssymb,
 aps,
]{revtex4-2}

\usepackage{cancel}
\usepackage{graphicx}
\usepackage{dcolumn}
\usepackage{bm}
\usepackage[
    colorlinks=true,       
    linkcolor=blue,        
    citecolor=blue,        
    urlcolor=blue,         
    pdftitle={Seu Título Aqui},
    pdfauthor={Seu Nome Aqui},
    linktocpage=false,      
    breaklinks=true,       
]{hyperref}

\newcommand{\mb}{\mathbf}

\begin{document}

\preprint{APS/123-QED}

\title{Topology of the Aharonov-Bohm effect in different reference frames}

\author{Hiram S. M. Rodrigues}\email{hiramrodrigues@ufmg.br}
 \affiliation{Departamento de F\'isica, Universidade Federal de Minas Gerais, Belo Horizonte, MG 31270-901, Brazil}
\author{Katson W. O. Arévola}\email{katsonw@ufmg.br}%
\affiliation{Departamento de F\'isica, Universidade Federal de Minas Gerais, Belo Horizonte, MG 31270-901, Brazil}%
\author{Pablo L. Saldanha}\email{saldanha@fisica.ufmg.br}\affiliation{Departamento de F\'isica, Universidade Federal de Minas Gerais, Belo Horizonte, MG 31270-901, Brazil}

\date{\today}

\begin{abstract}
Recent works showed that the Aharonov-Bohm (AB) phase difference for a quantum charged particle can be written in terms of electric and magnetic fluxes in a spacetime surface whose boundaries are the possible particle worldlines in the interferometer, relative to the possible paths. After presenting this result in a more detailed way, reinforcing its topological nature, we study the magnetic and electric versions of the AB effect in different inertial reference frames. We find a particular reference frame for a magnetic AB effect where the magnetic flux has a null contribution for the AB phase difference, which is entirely due to an electric flux. Also, we find a particular reference frame for an electric AB effect where the electric flux has a null contribution for the AB phase difference, which is entirely due to a magnetic flux. In this sense, the nomenclatures `magnetic AB effect' and `electric AB effect' lose their meaning. We have electromagnetic AB effects.
\end{abstract}

\maketitle


\section{Introduction}

As shown by the Aharonov-Bohm (AB) effect, given an interferometer whose paths enclose a region with a non-zero electromagnetic field, the interference pattern generated by  quantum charged particles depends on the existence of electromagnetic fields that do not interact locally with them \cite{ehrenberg49,aharonov59}. Several hypotheses have emerged to physically interpret the AB effect. It could be the result of a local interaction between the quantum particle and the electromagnetic potentials, implying that the potentials are physical entities rather than merely mathematical tools \cite{aharonov59}. Other possible interpretations suggest nonlocality emerging from the superposition of the particle field with the fields responsible for the potentials \cite{peshkin81,kang13,saldanha16} or from the interaction of the particle field with the source of the potentials \cite{vaidman12,pearle17a,pearle17b}, which happens away from the quantum particle position. More recently, the effect was described as the result of an exchange of virtual photons between the quantum charged particle and the source of the potentials, where both the particle and the source interact locally with the quantized electromagnetic field \cite{santos99,marletto20,saldanha21a,saldanha21b,saldanha24}. Independently from its interpretation, the AB effect was experimentally observed in many different systems \cite{chambers60,matteucci85,webb85,tonomura86,oudenaarden98,bachtold99,ji03,peng10,becker19,nakamura19,deprez21,ronen21}, being an example of a geometric and topological phase \cite{berry84,peshkin95,cohen19,ballentine}.

There are different nomenclatures for the AB effect in the literature. If the region enclosed by the interferometer paths has only a magnetic field, it is usually called `{magnetic} AB effect'. If this region has only an electric field, it is usually called `{electric} AB effect'. Recently, a scheme with time-varying electromagnetic fields was proposed, with the current in a solenoid varying while the quantum particle is in a superposition state inside two Faraday cages \cite{saldanha23,saldanha25}, which was named `{electrodynamic} AB effect'. By using a covariant formalism, the AB phase difference between the paths in any case can be written as a combination of two fluxes in spacetime: a magnetic flux and an electric flux through a spacetime surface whose boundaries are the two possible worldlines of the particle in the interferometer \cite{singleton13,saldanha23}. This description in terms of electromagnetic fluxes evidences the  topology of the AB effect in spacetime \cite{saldanha23,saldanha25}.

Considering the formalism in terms of electromagnetic fluxes of Refs. \cite{saldanha23,singleton13}, here we analyze the AB phase in different inertial reference frames, clarifying important aspects of its topological nature in spacetime. It is well known that a pure magnetic field in one reference frame may appear as a combination of electric and magnetic fields in another moving frame, and the same for a pure electric field. Following this reasoning, we show that the total AB phase difference is invariant under Lorentz transformations, but that the contributions of the electric flux and of the magnetic flux for the AB phase difference are frame-dependent. Somewhat surprisingly, for a magnetic AB effect with a particular interferometer geometry, we find a particular reference frame and a particular spacetime surface where there is no contribution of a magnetic flux for the AB phase difference, which is entirely due to an electric flux. Similarly, in an example of an electric AB effect we find a reference frame and a spacetime surface in which there is no contribution of an electric flux for the AB phase, which is entirely due to a magnetic flux. In this sense, as in the classical realm, the nomenclatures `electric AB effect' and `magnetic AB effect' may be misleading, since we actually  have electromagnetic AB effects.

This paper is divided in the following way: In section \ref{sec: The covariant Aharonov-Bohm phase} we obtain a form for the AB phase difference written in terms of electromagnetic fluxes, in a more comprehensive treatment than the ones from previous works \cite{singleton13,saldanha23}. In sections \ref{sec: Magnetic AB effect} and \ref{sec: Electric AB effect}, we consider the magnetic and electric versions of the AB effect in different reference frames, respectively. Finally, in section \ref{sec: Conclusion}, we present our concluding remarks.

\section{\label{sec: The covariant Aharonov-Bohm phase} Topology and covariance of the AB effect}

\begin{figure}
\includegraphics[scale=0.28]{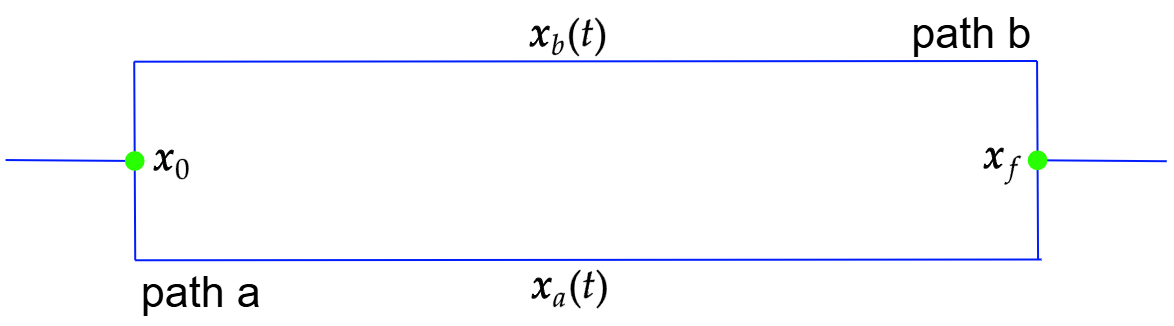}
\caption{\label{fig:ilustracao} General AB interferometer. The quantum charged particle has two possible paths in the interferometer, labeled $a$ and $b$.  $\bm{x}_{0}$ is the position where the incident particle wave function is split for both paths and $\bm{x}_{f}$ the position where the wave functions are recombined. The quantum particle always propagate in regions with null electromagnetic fields, but that may have nonzero potentials, such that we may have a nonzero AB phase difference between the paths.}
\end{figure}

In this section, we show how the Aharonov-Bohm phase difference can be expressed in terms of either the potentials or the fields via Stokes' theorem in 4D, describing the treatment from Ref. \cite{singleton13} in more detail. Consider a quantum particle with mass $m$ and charge $q$ that propagates along the arms of an interferometer, as in Fig. \ref{fig:ilustracao}, interacting with the electromagnetic potentials through the following interaction Hamiltonian \cite{ballentine}:
\begin{equation}
    H_I=qV-\frac{q}{m}\bm{p}\cdot\bm{A},
\end{equation}
\noindent where $\bm{p}$ is the particle momentum and $V$ and $\bm{A}$ are the scalar and vector potentials, respectively. It is assumed that the electromagnetic field at the possible particle positions is always null. Even so, the wave functions at each path may accumulate a phase, resulting in a phase difference at the end. We assume a non-relativistic regime in a small interferometer, such that the fields propagation times within the interferometer can be disregarded.

Assuming that the particle wave function is split at the position $\mb{x}_0$ at time $t=0$ in Fig. \ref{fig:ilustracao}, and that the wave packets can be localized by position vectors $\mb{x}_i(t)$ along each path, with $i=\{a,b\}$ (paths $a$ and $b$), the AB phase for each path is given by
\begin{equation}\label{phasei}
    \phi_i=-\frac{1}{\hbar}\int_{t_0}^{t_f}H_Idt=-\frac{q}{\hbar}\bigg(\int_{t_0}^{t_f}V_i\,dt-\int_{\bm{x}_0}^{\bm{x}_i}\bm{A}\cdot d\bm{x}\bigg).
\end{equation}
To derive the expression (\ref{phasei}), we performed the following semi-classical argument: writing $\bm{p}=md\bm{x}/dt$ implies that $(\bm{p}\cdot\bm{A}\,dt)/m=\bm{A}\cdot d\bm{x}$. We are thus considering wave-packets with reasonably well defined positions and momenta in paths $a$ and $b$, respecting the uncertainty relations.

Considering that the wave packets meet and interfere at position $\mb{x}_f$ at time $t_f$ in the scheme of Fig. \ref{fig:ilustracao}, the phase difference $\phi_{AB}\equiv\phi_a-\phi_b$ can be written as 
\begin{eqnarray}\label{faseABpotencial-tensor}
    \phi_{AB}=&&\frac{q}{\hbar}\bigg[-\underset{\text{path a}}{\int_{t_0}^{t_f}}V_a\,dt+\underset{\text{path b}}{\int_{t_0}^{t_f}}V_b\,dt+\nonumber \\
		&&\;\;\;\;\;\;+\underset{\text{path a}}{\int_{\bm{x}_0}^{\bm{x}_f}}\bm{A}\cdot d\bm{x} 
    -\underset{\text{path b}}{\int_{\bm{x}_0}^{\bm{x}_f}}\bm{A}\cdot d\bm{x}\bigg]  \nonumber\\
    =&&-\frac{q}{\hbar}\bigg[\oint \frac{V}{c}\,cdt+\oint A_i\,dx^i\bigg] =-\frac{q}{\hbar}\oint A_\mu dx^\mu,   
\end{eqnarray}
where the integration path in the last expressions for $\phi_{AB}$ goes forward through path $a$ and backwards through path $b$ in spacetime, considering the four-potential $A_\mu=(V/c,-\bm{A})$ and the four-differential $dx^\mu=(cdt,d\bm{x})$ with the metric signature $(+,-,-,-)$.

The expression of Eq. (\ref{faseABpotencial-tensor}) is explicitly covariant, being written in terms of the electromagnetic potentials. Now we want to arrive at an expression for the AB phase difference in terms of the electromagnetic fields. Eq. (\ref{faseABpotencial-tensor}) can be written in terms of differential forms \cite{singleton13}. Indeed, recognizing that the integrand in (\ref{faseABpotencial-tensor}) is the 1-form $A= A_\mu dx^\mu$, we can rewrite $\phi_{AB}$ as:
\begin{equation}\label{faseABpotencial-forma}
    \phi_{AB}=-\frac{q}{\hbar}\int A.
\end{equation}
Stokes' theorem in the language of forms is \cite{ryder}
\begin{equation}\label{stokes}
    \int_{\partial \Omega}\omega_p=\int_{\Omega}d\omega_p,
\end{equation}
where $\omega$ is a $p$-form, $d\omega$ is the exterior derivative of $\omega$, $\Omega$ is a $(p+1)$-chain, and $\partial\Omega$ is a $p$-chain that is the boundary of $\Omega$. In Appendix \ref{app:teorema de stokes} we present the traditional 3D versions of the divergence theorem and of Stokes' theorem in the language of forms, for those unfamiliar with this formalism.

\begin{figure}
\includegraphics[scale=0.17]{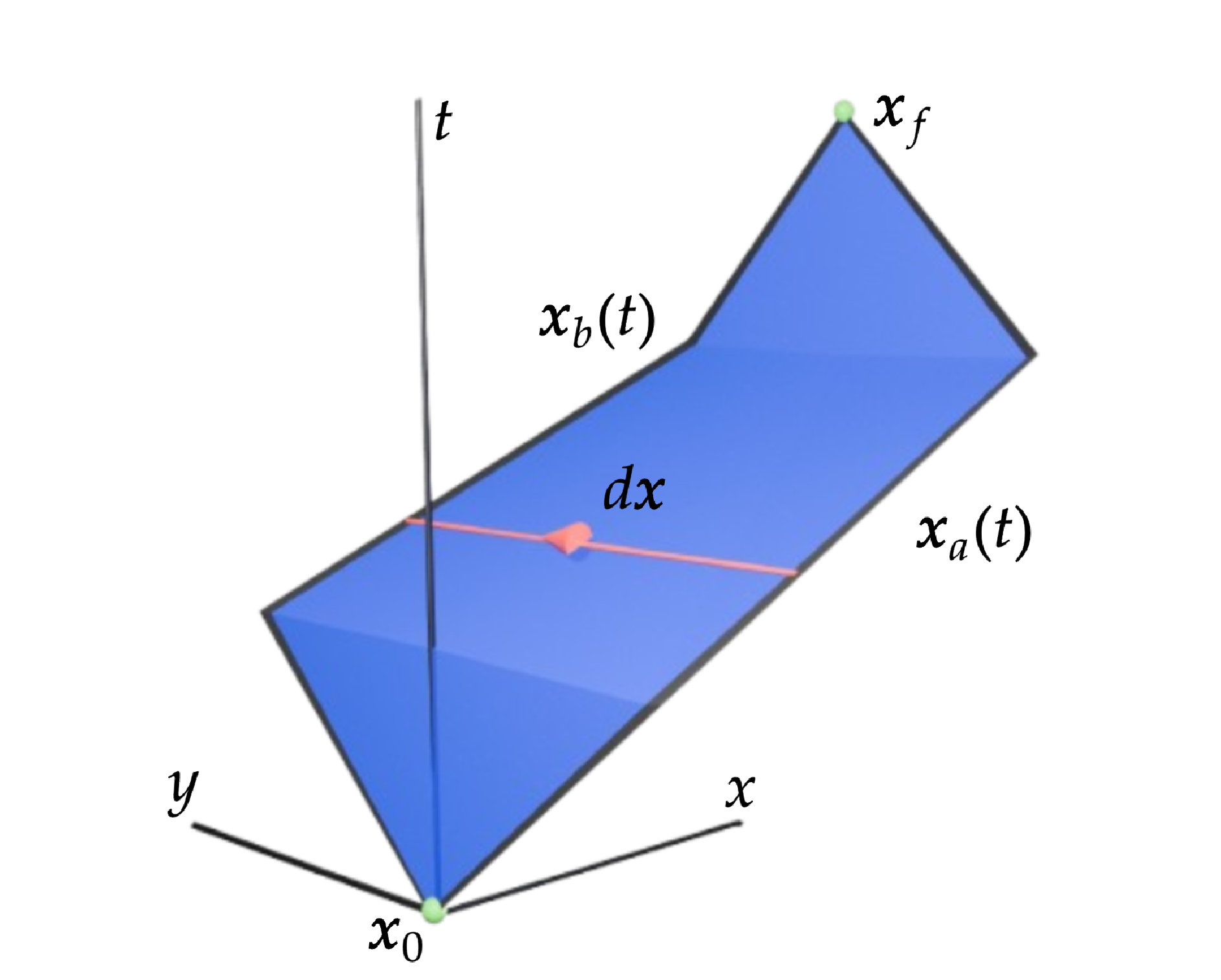}
\caption{\label{fig:ilustracaodeomega} Spacetime diagram of a quantum particle in the interferometer depicted in Fig. \ref{fig:ilustracao}. The interferometer is considered to be in the $xy$ plane and time is represented in the vertical direction. The two possible particle worldlines in the interferometer, $\mb{x}_a(t)$ and $\mb{x}_b(t)$, are represented in black. One possible spacetime surface $\Omega$ whose boundaries are the possible particle worldlines  is represented in blue. A curve connecting paths $a$ and $b$ through the surface $\Omega$ at a specific time is in red, with a differential displacement $d\bm{x}$ indicated. The origin of the system of coordinates was defined at $\bm{x}_{0}$. }
\end{figure}

Comparing Eqs. (\ref{faseABpotencial-forma}) and (\ref{stokes}), $A$ is a 1-form $\omega_1$ and the line integral in spacetime is represented by $\partial\Omega$ on the left side of Eq. (\ref{stokes}). On this way, $\Omega$ on the right side of Eq. (\ref{stokes}) is a surface in spacetime in this case, whose boundary is the closed curve $\partial\Omega$. So, as we wish to apply  Stokes' theorem from Eq. (\ref{stokes}) to Eq. (\ref{faseABpotencial-forma}), we must calculate $d\omega_1=dA$. Therefore, since $A=A_\mu dx^\mu=Vdt-\bm{A}\cdot d\bm{x}$,  using the property of the exterior product $dx^\mu\wedge dx^\nu=-dx^\nu\wedge dx^\mu$, we have
\begin{eqnarray}\label{aux}
       dA=&&\frac{\partial V}{\partial x}dx\wedge dt+\frac{\partial V}{\partial y}dy\wedge dt+\frac{\partial V}{\partial z}dz\wedge dt    \nonumber\\
    &&-\bigg[\frac{\partial A_x}{\partial t}dt\wedge dx+\frac{\partial A_y}{\partial t}dt\wedge dy+\frac{\partial A_z}{\partial t}dt\wedge dz \nonumber\\
    &&\;\;\;\;\;\;+\bigg(\frac{\partial A_y}{\partial x}-\frac{\partial A_x}{\partial y}\bigg)dx\wedge dy  \nonumber\\
    &&\;\;\;\;\;\;+\bigg(\frac{\partial A_z}{\partial y}-\frac{\partial A_y}{\partial z}\bigg)dy\wedge dz   \nonumber\\
    &&\;\;\;\;\;\;+\bigg(\frac{\partial A_x}{\partial z}-\frac{\partial A_z}{\partial x}\bigg)dz\wedge dx\bigg], \label{da-parcial}
\end{eqnarray}
 where $\wedge$ is the exterior product. The electromagnetic fields can be written as $\bm{E}=-\bm{\nabla}V-\partial_t\bm{A}$ and $\bm{B}=\bm{\nabla}\times\bm{A}$, such that Eq. (\ref{aux}) can be written as
\begin{equation}
    dA=\bm{E}\cdot(dt\wedge d\bm{x})-\bm{B}\cdot d\bm{a},\label{da}
\end{equation}
where $d\bm{a}\equiv(dy\wedge dz,dz\wedge dx,dx\wedge dy)$. On this way, the use of Eq. (\ref{stokes})  leads Eq. (\ref{faseABpotencial-forma}) to 
\begin{equation}  
    \phi_{AB}=\frac{q}{\hbar}\int_\Omega \{\bm{B}(\bm{x},t)\cdot d\bm{a}-\bm{E}(\bm{x},t)\cdot[dt d\bm{x}]\},  \label{faseABcampos-vetor} 
\end{equation}
where $d\bm{a}$ is the spatial component of the spacetime surface element at time $t$ and $d\bm{x}$ is the line element that goes from path $a$ to path $b$ through the spacetime surface at time $t$. The boundaries of the spacetime surface $\Omega$ are the two possible worldlines of the quantum particle in the interferometer, associated to the two different paths. From now on we will drop the symbol $\wedge$ from the equations. Figure \ref{fig:ilustracaodeomega} illustrates the possible particle worldlines and one possible spacetime surface $\Omega$ for the AB scheme depicted in Fig. \ref{fig:ilustracao}.

The AB phase of Eq. (\ref{faseABcampos-vetor}) is composed by two contributions, a magnetic flux and an electric flux  in spacetime \cite{saldanha23,singleton13}. In the magnetic AB effect, the region with a null magnetic field is not simply connected and the possible paths of the quantum particle in the interferometer enclose a region with nonzero magnetic field \cite{ehrenberg49,aharonov59}. The AB phase difference depends on the magnetic flux of Eq. (\ref{faseABcampos-vetor}), which do not depend on the specific geometries of the paths, only if they enclose or not the region with a nonzero magnetic field. So, the magnetic AB effect is a topological effect, depending on the topology of the possible particle paths relative to the magnetic field configuration \cite{berry84,peshkin95,cohen19,ballentine}. In the electric AB effect, the possible paths of the quantum particle in the interferometer enclose a region with nonzero electric field in spacetime, with the region with a null electric field not being simply connected in spacetime \cite{saldanha23}. On this way, the same topological arguments apply, and the electric AB effect is also a topological effect, with the AB phase difference depending on the electric flux of Eq. (\ref{faseABcampos-vetor}). The electrodynamic AB effect is also topological \cite{saldanha23,saldanha25}, with the AB phase difference depending on the electric and magnetic fluxes of Eq.  (\ref{faseABcampos-vetor}). Actually, Eq. (\ref{faseABcampos-vetor}), with the AB phase difference written in terms of the electromagnetic flux in a spacetime surface whose boundaries are the possible particle worldlines in the interferometer, evidences the topological nature of all electromagnetic AB effects. The examples treated in the next sections illustrate this point.

Different spacetime surfaces with the same boundaries may result in different contributions of the magnetic and electric fluxes in Eq. (\ref{faseABcampos-vetor}), generating the same AB phase difference \cite{saldanha23}. In the next sections we show how the contributions of the magnetic and electric fluxes of different AB schemes depend on the  reference frame, generating always the same AB phase difference.

\section{\label{sec: Magnetic AB effect} Magnetic AB effect in different reference frames}

\begin{figure}
\includegraphics[scale=0.16]{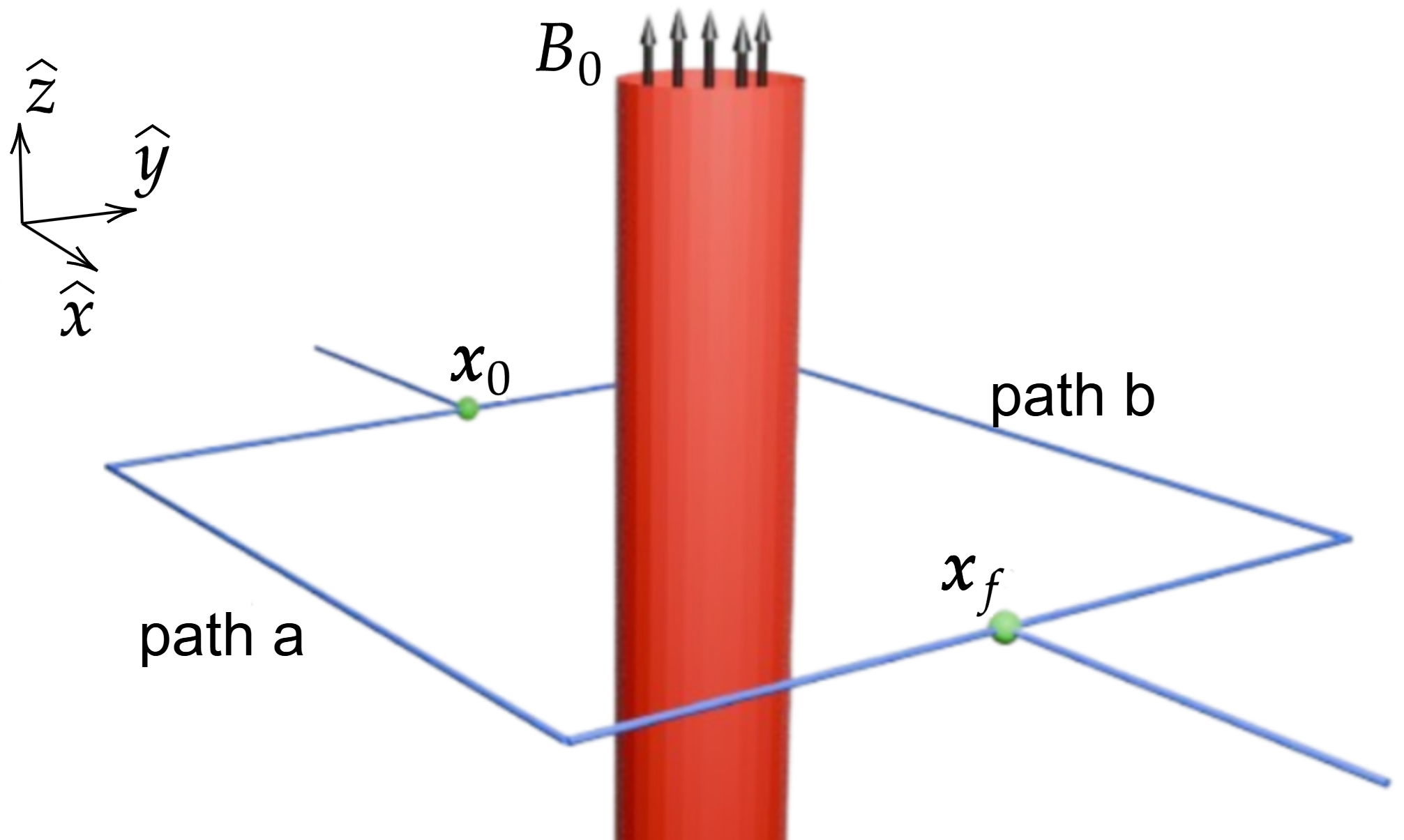}
\caption{\label{fig:casomag} Magnetic AB scheme with a square interferometer. The interferometer paths, in blue, enclose an infinite cylindrical solenoid, in red, which produces a uniform magnetic field inside it and a null field outside.}
\end{figure}

In this section, we explore the topological properties of the magnetic AB effect in different reference frames, illustrating interesting consequences. Consider the scheme shown in Fig. \ref{fig:casomag}, a square interferometer enclosing an infinite cylindrical solenoid with radius $r$, which produces a null magnetic field outside and a uniform field $\bm{B}_0$ inside it. 

In this case, in the interferometer rest frame $S$ the electric field is null and only the magnetic flux contributes to the AB phase from Eq. (\ref{faseABcampos-vetor}), which results in
\begin{eqnarray}
    \phi_{AB}^{S}=\frac{qB_o}{\hbar} \pi r^2. \label{faseABmagneticoS}
\end{eqnarray}

\noindent where $\pi r^2$ is the area of the solenoid cross section. 

Suppose now a reference frame $S'$, which is in relative motion with respect to frame $S$ along the $x$-axis with velocity $\bm{v}$. In the new frame, Eq. (\ref{faseABcampos-vetor}) is written as
\begin{equation}\label{faseABmagneticoS'}
    \phi_{AB}^{S'}=\frac{q}{\hbar}\int_{\Omega'} \{\bm{B'}(\bm{x'},t')\cdot d\bm{a'}-\bm{E'}(\bm{x'},t')\cdot[dt' d\bm{x'}]\},
\end{equation}

\noindent with the electromagnetic fields and the spatial and temporal components in frame $S'$ given by \cite{jackson,griffiths}
\begin{eqnarray}\label{transfolorentz}\nonumber
	&&\bm{E_\parallel'} =\bm{E_\parallel} \;,\;\;  \bm{E_\perp'}=\gamma\Bigl(\bm{E_\perp}+\bm{v}\times\bm{B}\Bigr), \quad \nonumber\\
	&&\bm{B_\parallel'}  = \bm{B_\parallel} \;,\;\; \bm{B_\perp'}= \gamma\Bigl(\bm{B_\perp}-\frac{\bm{v}}{c^2}\times\bm{E}\Bigr),\\\nonumber
	&&t' = \gamma\Bigl(t-\frac{v\,x}{c^2}\Bigr)\;,\;\; x'=\gamma(x-vt) \;,\;\;y' = y\;,\;\; z'= z,
\end{eqnarray}

\noindent where $\gamma=1/\sqrt{1-v^2/c^2}$, $\bm{E_\parallel}$ represents the component of the electric field in $S$ parallel to $\bm{v}$, $\bm{E_\perp}$ represents the component of the electric field in $S$ perpendicular to $\bm{v}$, and the same notation for the other fields. Using Eqs. (\ref{transfolorentz}) in Eq. (\ref{faseABmagneticoS'}), we obtain
\begin{equation}
   \int_{\Omega'}\bm{B'}(\bm{x'},t')\cdot d\bm{a'}=\gamma^{2}B_{0}\bigg[\iint_{\Omega} dxdy, 
   -\iint_{\Omega} v\,dtdy\bigg], \label{calculomag1}
\end{equation}
\begin{eqnarray}
    -\int_{\Omega'}\bm{E'}(\bm{x'},t')\cdot (dt'd\bm{x'})=&&\gamma^{2}B_{0}\bigg[\iint_{\Omega} v\,dtdy  \nonumber\\
    &&-\frac{v^2}{c^2}\iint_{\Omega}dxdy\bigg]. \label{calculoele1}
\end{eqnarray}

\noindent We thus see that in general we have a nonzero contribution of the electric flux for the AB phase of the magnetic AB effect in the frame $S'$, with a corresponding change of the magnetic flux that keeps the AB phase difference, obtained from Eqs. (\ref{faseABmagneticoS'}), (\ref{calculomag1}), and  (\ref{calculoele1}), the same, as it must be:
\begin{equation}\label{conclusaomag}
    \phi_{AB}^{S'}=\frac{q}{\hbar}B_{0}\gamma^{2}\underbrace{\bigg(1-\frac{v^2}{c^2}\bigg)}_{=\gamma^{-2}}\iint_{\Omega} dxdy=\frac{qB_0}{\hbar}\pi r^2=\phi_{AB}^{S}. 
\end{equation}

\begin{figure}
{\includegraphics[scale=0.18]{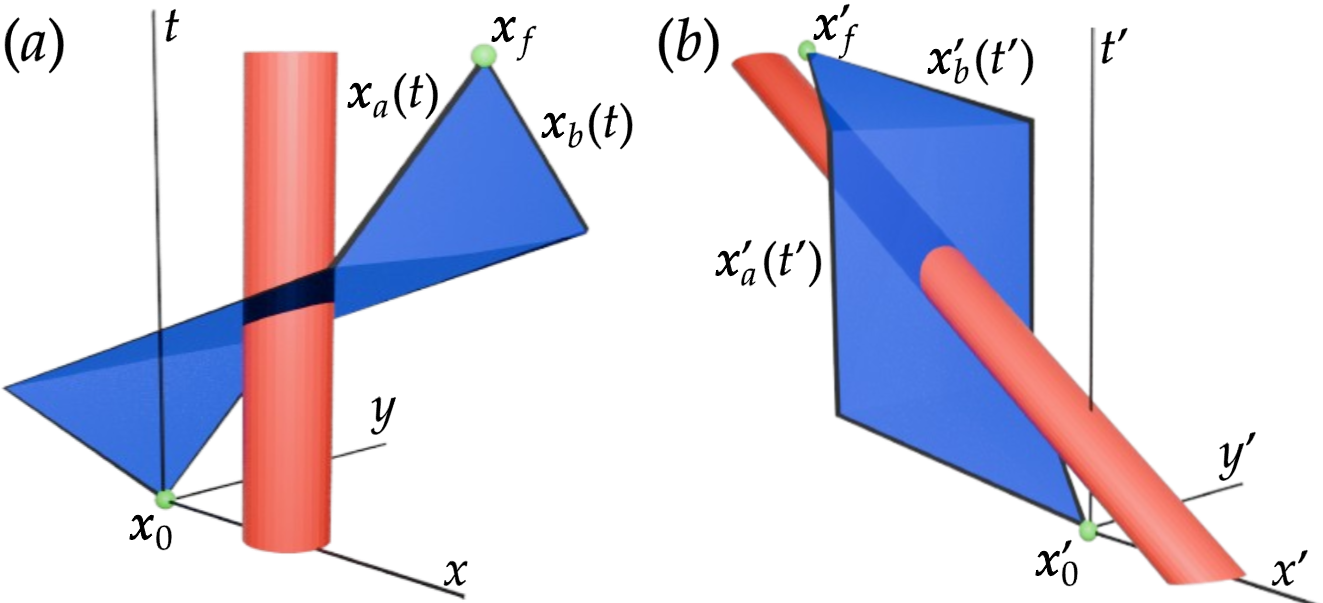}}
 \caption{\label{fig:espacotempo} Spacetime diagrams for the magnetic AB scheme of Fig. \ref{fig:casomag} in different reference frames, with the same representations as figure \ref{fig:ilustracaodeomega}. In both reference frames the solenoid position in the $xy$ plane is represented in red. (a) In the interferometer rest frame $S$, there is no electric field and the magnetic flux is the responsible for the AB phase of Eq. (\ref{faseABcampos-vetor}). (b) In this special reference frame $S'$, the particle velocity is null in the regions where it was propagating in the $x$ direction in frame $S$. On this way, the magnetic flux in the represented spacetime surface $\Omega'$ is null, since the area element is null in the superposition of this spacetime surface with the region where the magnetic field is nonzero, such that the electric flux is the responsible for the AB phase of Eq. (\ref{faseABmagneticoS'}).}
\end{figure}

Figure \ref{fig:espacotempo} depicts the spacetime diagrams of the magnetic AB effect of the scheme of Fig. \ref{fig:casomag} in two reference frames. Fig. \ref{fig:espacotempo}(a) shows the spacetime diagram in the interferometer rest frame $S$. Fig. \ref{fig:espacotempo}(b) represents the particular case where the magnitude of the relative velocity between $S'$ and $S$ is the same as the velocity of the wave packet in the interferometer rest frame. It can be seen in Fig. \ref{fig:espacotempo}(b) that the projection of the intersection between the solenoid region (where the field is nonzero) and the surface $\Omega'$ onto the spatial plane $x'y'$ results in line element, not a surface element. Consequently, the magnetic flux is null through this spacetime surface. Thus, we conclude that in the frame  $S'$ only the electric flux contribute to the AB phase in Eq. (\ref{faseABmagneticoS'}), as we show in the following.

Considering Eqs. (\ref{transfolorentz}), in the frame $S'$ we have a uniform electric field $\bm{E'}=E'_0\bm{\hat{y}}$ inside the solenoid and a null field outside it, with $E'_0=-\gamma vB_0$. For the surface $\Omega'$ represented in Fig. 
\ref{fig:espacotempo}(b), we can write $d\bm{x'}=dy \bm{\hat{y}}$ in the spatial integral of the electric flux in Eq. (\ref{faseABmagneticoS'}). On this way, the electric flux from Eq. (\ref{faseABmagneticoS'}) results in $E'_0$ times the `area' in spacetime of the ellipse in Fig. \ref{fig:espacotempo}(b) resultant from the intersection of the spacetime surface $\Omega'$ (in blue) with the solenoid region (in red), which has a nonzero electric field. The diameter of the ellipse in the spatial direction is $2r$, the diameter of the solenoid. Its `diameter' in the time direction, $\Delta t'$, is equal to the time that the solenoid takes to completely cross the line that connects the possible particle positions in each path (these positions are fixed in the superposition of the blue surface with the red region in Fig. \ref{fig:espacotempo}(b)). Due to Lorentz contraction, the width of the solenoid in the $x$ direction is $2r/\gamma$, while the solenoid velocity is $v$. In this way, we have $\Delta t'=2r/(v\gamma)$ and the AB phase of Eq. (\ref{faseABmagneticoS'}), due only to the electric flux, can be written as 
\begin{equation}\label{faseabS'-electricflux-magneticAB}
    \phi_{AB}^{S'}=-\frac{qE'_0}{\hbar}\frac{\pi \;2r \;\Delta t'}{4} = \frac{qB_0}{\hbar}\pi r^2 = \phi_{AB}^{S}.  
\end{equation}

In any reference frame, the region with a nonzero electromagnetic field is not simply connected in spacetime, as seen in Fig. \ref{fig:espacotempo}. The AB phase difference depends on the electromagnetic flux enclosed by the possible particle paths, as evidenced in Eqs. (\ref{faseABcampos-vetor}) or (\ref{faseABmagneticoS'}), not on details of the paths geometry, characterizing a topological effect.
 
The magnetic AB effect represented in Fig. \ref{fig:casomag} has this interesting peculiarity. It is called a \textit{magnetic} AB effect because in the laboratory frame $S$ we have only magnetic fields, the electric field being zero, such that the AB phase difference between the paths is due to the magnetic flux in Eq. (\ref{faseABcampos-vetor}). However, in the reference frame $S'$ the magnetic contribution for the AB phase difference in Eq. (\ref{faseABmagneticoS'}) is null for the spacetime surface $\Omega'$ represented in Fig. \ref{fig:espacotempo}(b), with the phase coming entirely from the electric flux. So, the nomenclature `magnetic AB effect' completely loses its meaning in the reference frame $S'$.

\section{\label{sec: Electric AB effect} Electric AB effect in different reference frames}

\begin{figure}
\includegraphics[scale=0.27]{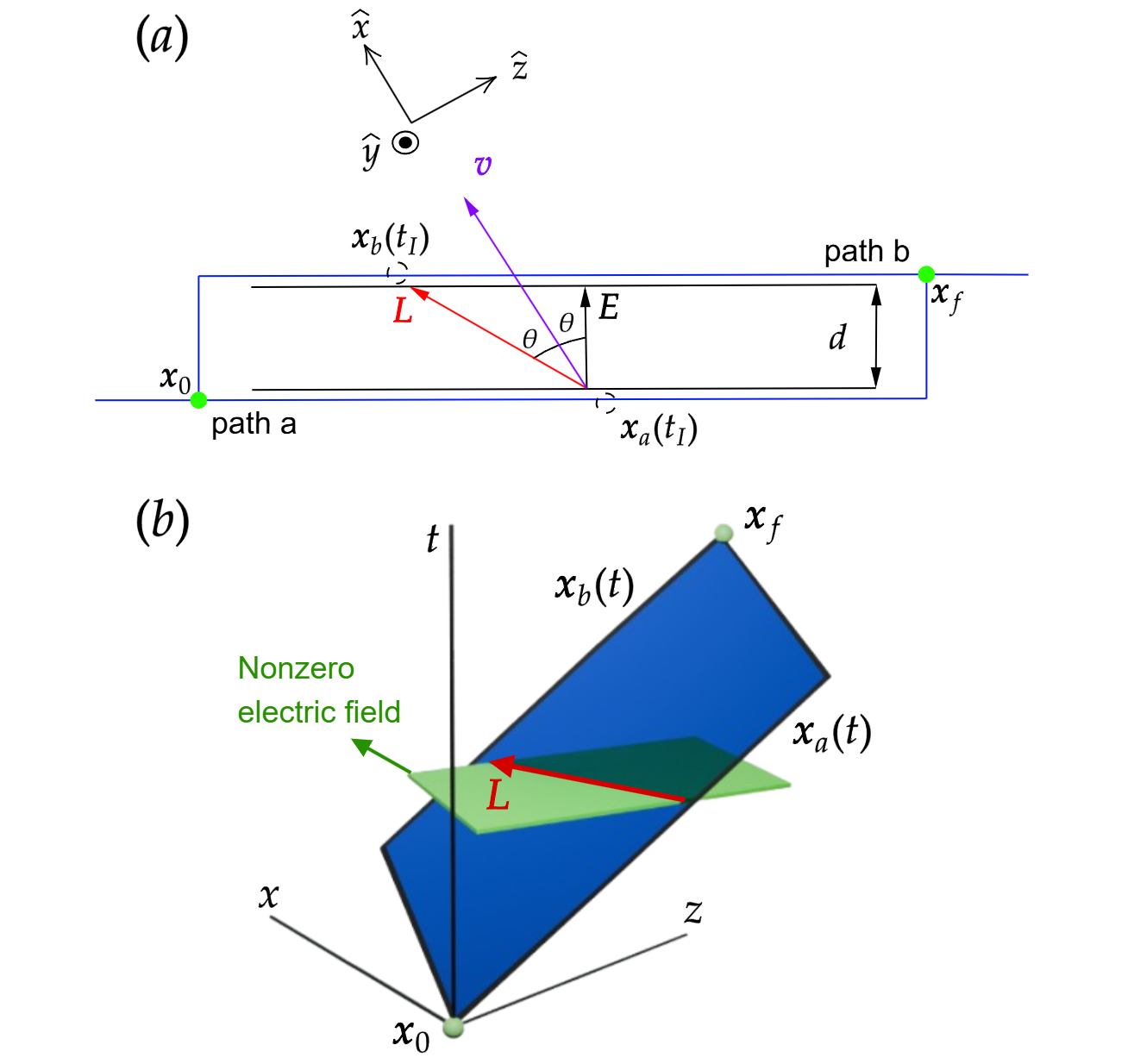}
\caption{\label{fig:casoele}
(a) Electric AB effect in the laboratory frame $S$. The potential difference between the interferometer paths  (in blue) is created by a parallel-plate capacitor (in red), which is charged at time $t_I$ producing a uniform electric field $\bm{E}$ inside it while the quantum particle is in a superposition state at the positions $\bm{x}_a(t_I)$ and $\bm{x}_b(t_I)$. The capacitor is charged for a short period $T$, such that the particle movement can be disregarded during this period. $\bm{L}$ represents the portion of the straight line that connects $\bm{x}_a(t_I)$ to $\bm{x}_b(t_I)$ inside the capacitor. The velocity $\bm{v}$ between the reference frame $S'$ and the interferometer rest frame $S$ is also represented. (b) Spacetime diagram of a quantum particle in the interferometer depicted in panel (a), with the same representations as figure \ref{fig:ilustracaodeomega}. The spacetime region with a nonzero electric field is represented in green. The vector $\bm{L}$ is represented as the red arrow.}
\end{figure}

In this section we consider interesting peculiarities of the electric AB effect in different reference frames, similar to the ones obtained for the magnetic AB effect in the previous section. Consider the scheme of Fig. \ref{fig:casoele}(a), with a parallel-plates capacitor between the possible trajectories of the quantum charged  particle. At time $t_I$, the capacitor is charged while the particle wave packets are away from the edges, superposed at the positions $\bm{x}_a(t_I)$ and $\bm{x}_b(t_I)$. The capacitor is charged only for a short period $T$, producing a uniform electric field $\bm{E}=E(\sin\theta\,\bm{\hat{z}}+\cos\theta\,\bm{\hat{x}})$ inside it, and the particle moves a negligible amount during this period. Disregarding the border effects, the electric field at the positions $\bm{x}_a(t_I)$ and $\bm{x}_b(t_I)$ is null, such that the particle always propagates in regions with a null field. As seen in the Fig. \ref{fig:casoele}(a), $\bm{L}$ is the portion of the straight line that connects $\bm{x}_a(t_I)$ to $\bm{x}_b(t_I)$ inside the capacitor, and we have $d = L\cos(2\theta)$.

Since in the interferometer rest frame $S$ the system magnetic field is zero, according to Eq. (\ref{faseABcampos-vetor}) only the electric flux contributes to the AB phase, which is given by
\begin{equation}\label{faseABeletS}
    \phi_{AB}^{S}=-\frac{qEdT}{\hbar}=-\frac{qELT\cos(2\theta)}{\hbar}. 
\end{equation}

\noindent Fig. \ref{fig:casoele}(b) shows the spacetime diagram of the described electric AB effect.

Now let us analyze the same scheme in a reference frame $S'$ that moves with velocity $\bm{v}$ in the $x$ direction. As depicted in Fig. \ref{fig:casoele}(a), $\bm{v}$ makes an angle $\theta$ with $\bm{E}$ and also an angle $\theta$ with $\bm{L}$. The coordinates and fields in the new frame are given by Eqs. (\ref{transfolorentz}) and the AB phase difference by Eq. (\ref{faseABmagneticoS'}). Due to Lorentz contraction, the component of $\bm{L}$ parallel to $\bm{v}$ is contracted by a factor $\gamma$ in the new frame, with the orthogonal component unaffected. And according to Eqs. (\ref{transfolorentz}), the component of $\bm{E}$ orthogonal to $\bm{v}$ is increased by a factor $\gamma$ in the new frame, with the parallel component unaffected. On this way, we have
\begin{equation}
	\mb{E}'\cdot\mb{L}'=\frac{E_\parallel L_\parallel}{\gamma}-\gamma E_\perp L_\perp=\frac{EL}{\gamma}(\cos^2\theta-\gamma^2\sin^2\theta)
\end{equation}
 
\noindent as the result of the spatial integral of the electric flux from Eq. (\ref{faseABmagneticoS'}). Therefore, under the condidion
\begin{equation}\label{condição fluxo nulo}
    \cos^2\theta=\gamma^2\sin^2\theta,
\end{equation}
we have $\mb{E}'\cdot\mb{L}'=0$ and the contribution of the electric flux to the AB phase of Eq. (\ref{faseABmagneticoS'}) is null in reference frame $S'$. As we show in the following, the AB phase in this frame is only due to the magnetic flux in Eq. (\ref{faseABmagneticoS'}).

\begin{figure}
\includegraphics[scale=0.20]{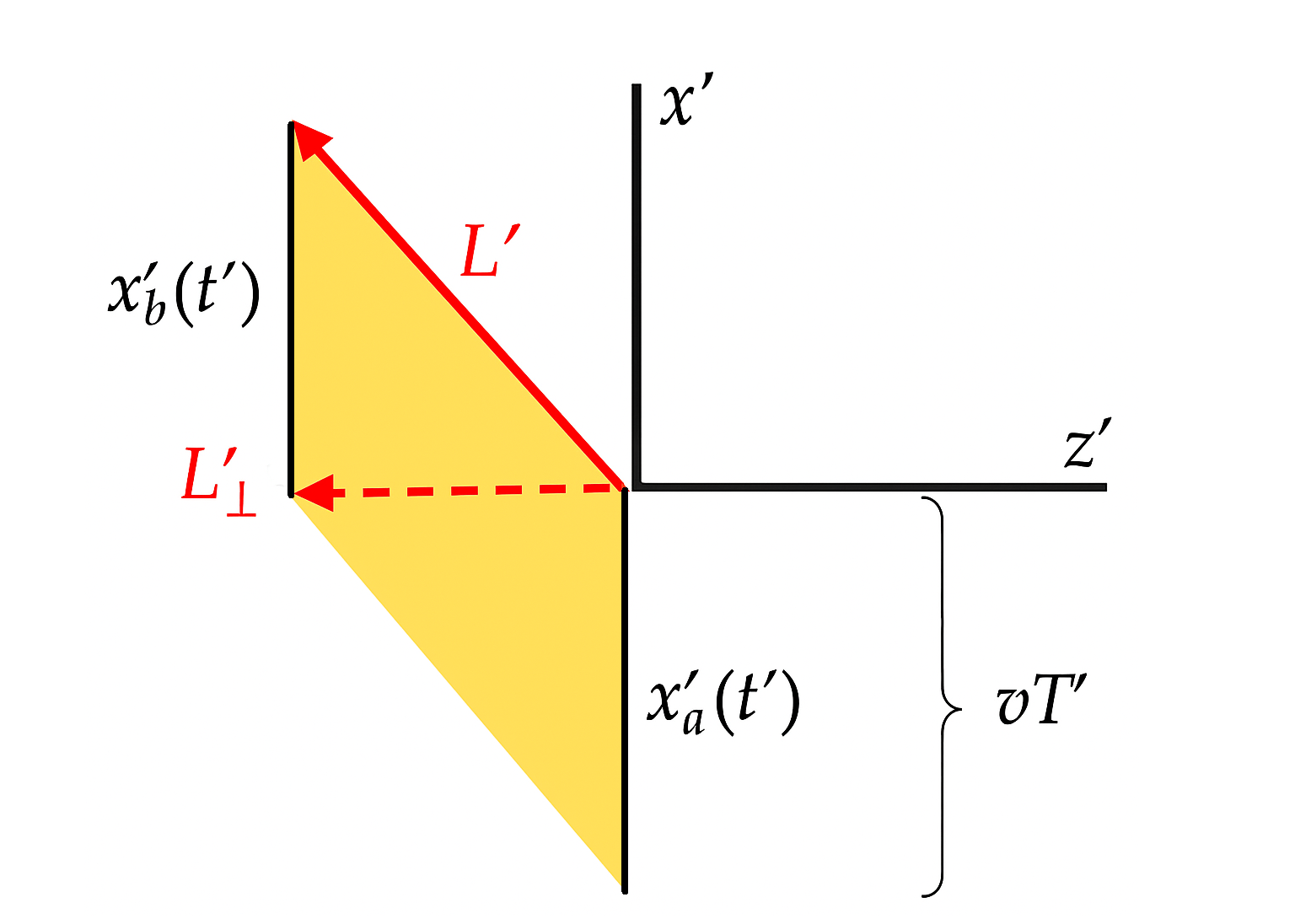}
\caption{\label{fig:planoomegaemS'casoele} In frame $S'$, the vector $\bm{L}$ moves with velocity $-v\bm{\hat{x}}$, spanning the area indicated in yellow while the field in the capacitor is nonzero.}
\end{figure}

In frame $S'$, according to Eq. (\ref{transfolorentz}) we have a magnetic field $\bm{B}'=\gamma vE\sin\theta/c^2\bm{\hat{y}}$ inside the capacitor. In this new frame, the vector $\bm{L}'$, which is the portion of the straight line that connects the positions of the superposed particle wave functions inside the capacitor,  moves with velocity $-v\bm{\hat{x}}$, spanning the area indicated by the blue surface in Fig. \ref{fig:planoomegaemS'casoele} while the field is turned on, during an interval $T'$. The area of the parallelogram in Fig. \ref{fig:planoomegaemS'casoele} is $vT'L'_\perp$, with $T'=\gamma T$ due to time dilatation and $L'_{\perp}=L_{\perp}=L\sin\theta$. On this way, the AB phase of Eq. (\ref{faseABmagneticoS'}), which is resultant only from the magnetic flux, is given by
\begin{equation}\label{eletS'}
	\phi_{AB}^{S'}=-\frac{qELT\gamma^2 v^2\sin^2\theta}{\hbar c^2}=-\frac{qELT}{\hbar}(\gamma^2-1)\sin^{2}\theta.
\end{equation}
Using the condition of Eq. (\ref{condição fluxo nulo}), it can be seen that $\phi_{AB}^{S'}$ from Eq. (\ref{eletS'}) is equal to $\phi_{AB}^{S}$ from Eq. (\ref{faseABeletS}). 

In any reference frame, the region with a nonzero electromagnetic field is not simply connected in spacetime, as seen in Fig. \ref{fig:casoele}(b) for the reference frame $S$. As evidenced by Eqs. (\ref{faseABcampos-vetor}) or (\ref{faseABmagneticoS'}), the AB phase difference depends on the electromagnetic flux enclosed by the possible particle paths, not on details of the paths geometry, characterizing a topological effect.

The electric AB effect treated in this section has this interesting behavior, similar to the magnetic AB effect treated in the previous section. It is called an \textit{electric} AB effect because in the laboratory frame $S$ we have only electric fields, the magnetic fields being zero, such that the AB phase difference between the paths is due to the electric flux in Eq. (\ref{faseABcampos-vetor}). However, in the reference frame $S'$ the electric contribution for the AB phase difference in Eq. (\ref{faseABmagneticoS'}) is null, with the phase coming entirely from the magnetic flux. So, the nomenclature `electric AB effect' completely loses its meaning in the reference frame $S'$.

\section{Conclusion}\label{sec: Conclusion}

The AB phase difference between the two paths a charged quantum particle may pass in an interferometer do not depend on the reference frame, as is evident in the relativistic notation of Eq. (\ref{faseABpotencial-tensor}). It can be written in terms of the electromagnetic potentials at the possible particle positions, as in Eq. (\ref{faseABpotencial-tensor}), or in terms of the electric and magnetic fluxes in a spacetime surface whose boundaries are the possible particle worldlines in the interferometer, as in Eq. (\ref{faseABcampos-vetor}) \cite{singleton13,saldanha23}. This last expression evidences the topological nature of the AB effect in spacetime.

The contributions of the electric and magnetic fluxes to the AB phase difference of Eq. (\ref{faseABcampos-vetor}) depend on the reference frame, and our work explored this dependence to present interesting consequences of this fact, evidencing somewhat surprising features of the topological nature of the AB effect. For instance, if an `electric AB effect' is defined as one in which we have only an electric flux contribution for the AB phase difference of Eq. (\ref{faseABcampos-vetor}), with a null contribution from the magnetic flux, and a `magnetic AB effect' as one in which we have only a magnetic flux contribution for the AB phase, we conclude that what is an electric AB effect in one reference frame can be seen as a magnetic AB effect in another frame, and vice-versa. Such conclusions evidence that we actually have \textit{electromagnetic} AB effects.

\begin{acknowledgments}

This work was supported by the Brazilian agencies CNPq (Conselho Nacional de Desenvolvimento Cient\'ifico e Tecnol\'ogico) and CAPES (Coordenação de Aperfeiçoamento de Pessoal de Nível Superior).

\end{acknowledgments}

\appendix

\section{\label{app:teorema de stokes}Divergence and Stokes' theorems in the language of forms}

Here we present a description of the divergence theorem and of Stokes' theorem in 3D in the language of forms from Eq. (\ref{stokes}), to introduce the subject for the readers unfamiliar with it.

The divergence theorem
\begin{equation}
	\oint_{S}\bm{A}\cdot d\bm{a}=\int_V\bm{\nabla}\cdot\bm{A}\,dV
\end{equation}
is just a particular case of (\ref{stokes}) for the 2-form $\omega_2=A_xdy\wedge dz+A_ydz\wedge dx+A_zdx\wedge dy$ integrated in a 2-chain surface $\partial \Omega=S$, which is the boundary of the 3-chain volume $\Omega=V$. On the right hand side we have the integral of the exterior derivative of $\omega_2$ in $V$, with
\begin{equation}
	d \omega_2=\frac{\partial A_x}{\partial x}dx\wedge dy\wedge dz+\frac{\partial A_y}{\partial y}dy\wedge dz\wedge dx+\frac{\partial A_z}{\partial z}dz\wedge dx\wedge dy.
\end{equation}
Note that the property of the exterior product $dx^\mu\wedge dx^\nu=-dx^\nu\wedge dx^\mu$ makes any term $dx^\mu\wedge dx^\mu$ null. 

The 3D version of Stokes' theorem 
\begin{equation}
	\oint_{C}\bm{A}\cdot d\bm{x}=\int_S(\bm{\nabla\times\bm{A}})\cdot d\bm{a}
\end{equation}
is another a particular case of (\ref{stokes}) for the 1-form $\omega_1=A_xdx+A_ydy+A_zdz$ integrated in a 1-chain curve $\partial \Omega=C$, which is the boundary of the 2-chain surface $\Omega=S$. On the right hand side we have the integral of the exterior derivative of $\omega_1$ in $S$, with
\begin{eqnarray}
    d\omega_1&&=\frac{\partial A_x}{\partial y}dy\wedge dx+\frac{\partial A_x}{\partial z}dz\wedge dx+\frac{\partial A_y}{\partial x}dx\wedge dy  \nonumber\\
    &&+\frac{\partial A_y}{\partial z}dz\wedge dy+\frac{\partial A_z}{\partial x}dx\wedge dz+\frac{\partial A_z}{\partial y}dy\wedge dz.
\end{eqnarray}


%

\end{document}